# Robust Frame and Frequency Synchronization Based on Alamouti Coding for RGI-CO-OFDM

Oluyemi Omomukuyo, Deyuan Chang, Octavia Dobre, Ramachandran Venkatesan, and Telex M. N. Ngatched

***Abstract*— We propose an algorithm for carrying out joint frame and frequency synchronization in reduced-guard-interval coherent optical orthogonal frequency division multiplexing (RGI-CO-OFDM) systems. The synchronization is achieved by using the same training symbols (TS) employed for training-aided channel estimation (TA-CE), thereby avoiding additional training overhead. The proposed algorithm is designed for polarization division multiplexing (PDM) RGI-CO-OFDM systems that use the Alamouti-type polarization-time coding for TA-CE. Due to their optimal TA-CE performance, Golay complementary sequences have been used as the TS in the proposed algorithm. The frame synchronization is accomplished by exploiting the cross-correlation between the received TS from the two orthogonal polarizations. The arrangement of the TS is also used to estimate the carrier frequency offset. Simulation results of a PDM RGI-CO-OFDM system operating at 238.1 Gb/s data rate (197.6-Gb/s after coding), with a total overhead of 9.2% (31.6% after coding), show that the proposed scheme has accurate synchronization, and is robust to linear fiber impairments.**

## I. Introduction

COHERENT optical orthogonal frequency division multiplexing (CO-OFDM) has several advantages including high spectral efficiency, and robustness to both chromatic dispersion (CD) and polarization mode dispersion (PMD) [1]. Polarization division multiplexing (PDM) is an effective technique for doubling the spectral efficiency of CO-OFDM systems. For long-haul transmission, in order to avoid using a long length of cyclic prefix (CP) for CD compensation, reduced-guard-interval CO-OFDM (RGI-CO-OFDM), which employs a short CP length has been proposed [2]. It is well-known that PDM optical communication systems are analogous to 2×2 multiple-input multiple-output (MIMO) wireless systems. Consequently, space-time block code (STBC) algorithms [3] can be readily applied in PDM coherent optical systems for training-aided channel estimation (TA-CE). In this context, the STBC codes are referred to, more appropriately, as polarization-time (PT) codes. PT codes have also been applied in PDM systems for other applications, including PMD compensation [4], [5], and mitigating polarization-dependent loss (PDL) [6]. Recently, PT codes have also been proposed for training-aided joint frame and frequency synchronization in single-carrier PDM systems [7].

For PDM CO-OFDM systems, several joint schemes have also been proposed for training-aided frame and frequency synchronization, see e.g., [8]. However, to the best of our knowledge, these schemes require extra overheads because they do not use the same training symbols (TS) to carry out both the joint synchronization and TA-CE. In addition, these schemes may be limited by one or more of the following: frame synchronization errors under poor optical signal-to-noise ratio (OSNR) conditions, limited carrier frequency offset (CFO) estimation range, and complex TS structures. We previously proposed a joint synchronization algorithm for a single-polarization RGI-CO-OFDM system [9], which used only one simple-structured TS for synchronization, and demonstrated better performance when compared with popular existing synchronization algorithms. Although the method in [9] can be readily extended for PDM transmission, it would still be disadvantaged by the need for additional TS for TA-CE.

In this letter, to overcome the above-mentioned limitations, we propose and demonstrate the use of Alamouti-type [3] PT codes for joint frame and frequency synchronization in a 238.1-Gb/s 16-ary quadrature amplitude modulation (16-QAM) PDM RGI-CO-OFDM system. To ensure that optimal TA-CE performance [7] is achieved, the TS used for both the joint synchronization and TA-CE are Golay complementary sequences (GCS) [10]. The proposed scheme has a wide CFO estimation range, and is robust to linear fiber impairments.

## II. Operation Principle

The GCS used as the TS in the proposed scheme satisfy the autocorrelative property [10]:

$$\sum_{m=0}^{L-1-j} a_m a_{m+j}^* + \sum_{m=0}^{L-1-j} b_m b_{m+j}^* = \begin{cases} 0, & j \neq 0 \\ 2L, & j = 0 \end{cases}, \qquad (1)$$

(a)

| | | |
|---|---|---|
| X-Pol | A(m) | -B(m)* |
| Y-Pol | B(m) | A(m)* |

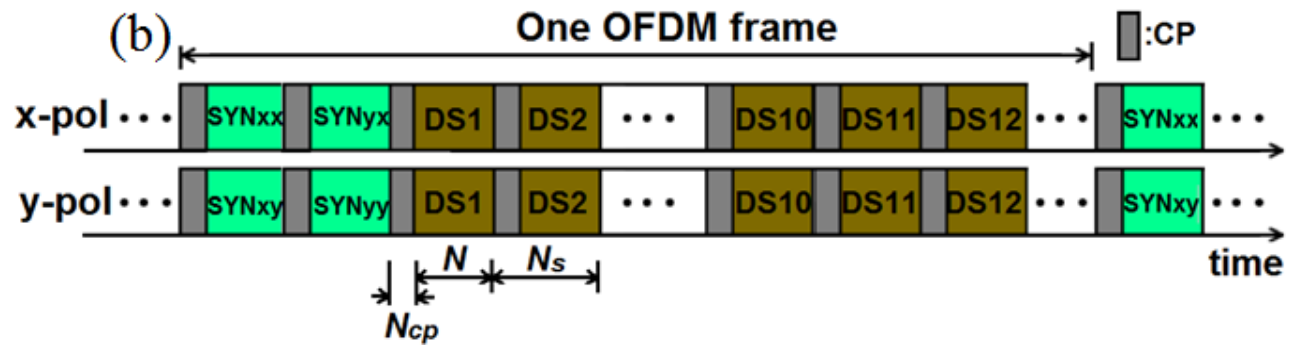

Fig. 1. (a) Training symbol arrangement, and (b) frame structure for the PDM RGI-CO-OFDM system. SYN: synchronization symbol. DS: data symbol.

where the superscript * is the complex conjugation operator, $j = 0,1,2,\cdots,L-1$, and $a_m$, $b_m$ $(m = 0,1,2,\cdots,L-1)$ are elements of the GCS, $A(m)$ and $B(m)$, of length $L$. The TS are arranged in the Alamouti scheme as shown in Fig. 1(a). Since the CO-OFDM system would typically require unmodulated subcarriers for purposes such as oversampling, $L$ is not chosen to be a power of 2 in this work, but is made equal to the number of data subcarriers. This would enable the TS to have the same mapping structure to an $N$-point inverse Fast Fourier Transform (IFFT) as the rest of the data symbols. To achieve this, a binary Golay seed pair of length 26 [11] is first mapped to QPSK symbols by making use of Gray mapping. Quaternary GCS of length 416 are then obtained recursively from the QPSK symbols. Finally, the quaternary GCS are converted to 16-QAM GCS using the method in [12].

The TS and data symbols are input to the IFFT, yielding a time-domain frame as shown in Fig. 1(b). Each symbol in the frame has a length of $N_s = N + N_{cp}$ samples, where $N_{cp}$ is the CP duration. In Fig. 1(b), $SYN_{yx}$ and $SYN_{yy}$ represent the IFFTs of $-B(m)^*$ and $A(m)^*$, respectively, while $SYN_{xx}$ and $SYN_{xy}$ represent the IFFTs of $A(m)$ and $B(m)$, respectively, multiplied by the corresponding elements of a real-valued pseudo-random (PN) sequence, $p(n) \in (-1,1]$, with $n = 0,1,\cdots,N_s-1$.

### A. Frame Synchronization

The frame synchronization procedure relates to the algorithm for identifying Alamouti STBC OFDM signals in wireless MIMO [13]. By exploiting the complementary property of the GCS, together with the Alamouti scheme arrangement, we can define timing metrics as follows:

$$M_{x(y)}(d) = \frac{\left|P_{x(y)}(d)\right|^2}{R^2_{x(y)}(d)}, \tag{2}$$

$$P_{x(y)}(d) = \sum_{n=0}^{N_s-1}\left(P^A_{x(y)}(d) - P^B_{x(y)}(d)\right), \tag{3}$$

$$P^A_x(d) = r_x(d+n)p(n) \times r_y\left(d + \alpha + N_r + mod\left(\left[N_{cp}-n\right],N\right)\right), \tag{4}$$

$$P^B_x(d) = r_y(d+\alpha+n)p(n) \times r_x\left(d + N_r + mod\left(\left[N_{cp}-n\right],N\right)\right), \tag{5}$$

$$P^A_y(d) = r_x(d-\alpha+n)p(n) \times r_y\left(d + N_r + mod\left(\left[N_{cp}-n\right],N\right)\right), \tag{6}$$

$$P^B_y(d) = r_y(d+n)p(n) \times r_x\left(d - \alpha + N_r + mod\left(\left[N_{cp}-n\right],N\right)\right), \tag{7}$$

$$R_{x(y)}(d) = 2\sum_{n=0}^{N_s-1}\left|r_{x(y)}(d+n)\right|^2, \tag{8}$$

where $N_r = N_s + N_{cp}$, $r_x(d)$ and $r_y(d)$ are the discrete received samples for the $x$ and $y$ polarizations respectively, $d$ is the time index corresponding to the first received sample in a window of $N_s$ samples, $mod(.)$ is the modulo operator, $\alpha$ is the relative delay (in samples) between the two polarizations caused by PMD, where $\alpha = -\beta, -\beta+1, \cdots, \beta$, and $\beta$ is the maximum relative delay. If $p(n)$ is not included in (4)-(7), the timing metric for each polarization will have a plateau with a width equal to $\beta$, which can cause frame synchronization errors. In addition, to avoid these errors, it is necessary to have $\beta \geq \lceil DGD \times F_s \rceil$, where $\lceil . \rceil$ is the ceiling operator, DGD is the dispersion group delay, and $F_s$ is the sampling rate. The estimate of the start of the frame for each polarization, $\hat{d}_{x(y)}$, is obtained as the index that maximizes (2).

### B. Frequency Synchronization

The CFO estimate, $\hat{\gamma}$, is obtained as:

$$\hat{\gamma} = (\hat{\varepsilon} + \hat{\mu})\Delta f_N, \tag{9}$$

where $\hat{\varepsilon}$ and $\hat{\mu}$ are the estimates of the fractional and integer parts of the normalized CFO respectively, and $\Delta f_N$ is the OFDM subcarrier frequency spacing. First, $\hat{\varepsilon}$ is obtained as:

$$\hat{\varepsilon} = \frac{1}{\pi} arg\left[\sum_{n=0}^{\frac{N_s}{2}-1}\left(P^A_{x(y)}\left(\hat{d}_{x(y)}\right) - P^B_{x(y)}\left(\hat{d}_{x(y)}\right)\right)\right], \tag{10}$$

where $P^A_{x(y)}\left(\hat{d}_{x(y)}\right)$ and $P^B_{x(y)}\left(\hat{d}_{x(y)}\right)$ are obtained from (4) and (5) by setting $\alpha = 0$ and $d = \hat{d}_{x(y)}$. After fractional CFO compensation, $\hat{\mu}$ is obtained as the index that maximizes

$$\Xi(\mu) = \frac{\left|\sum_{v=0}^{N-1}\left(A_f + B_f\right)^*(v)R_f(v+\mu)\right|^2}{\left(\sum_{v=0}^{N-1}\left|R_f(v+\mu)\right|^2\right)^2}, \tag{11}$$

with $\mu = -\frac{N}{2}, -\frac{N}{2}+1, \cdots, \frac{N}{2}-1$, where $A_f(v)$ and $B_f(v)$ are obtained by padding $A(m)$ and $B(m)$ with $N-L$ zeros, respectively, and $R_f(v)$ is obtained by first multiplying the received samples of $SYN_{xx}$ by the corresponding elements of $p(n)$, and then carrying out an $N$-point FFT of this product. Given that $|\hat{\varepsilon}| \leq 1$, it can be deduced from (9)-(11) that the theoretical CFO estimation range of the proposed algorithm is $-\frac{N\Delta f_N}{2} \leq \hat{\gamma} \leq \frac{N\Delta f_N}{2}$, which is as wide as $\pm\frac{F_s}{2}$.

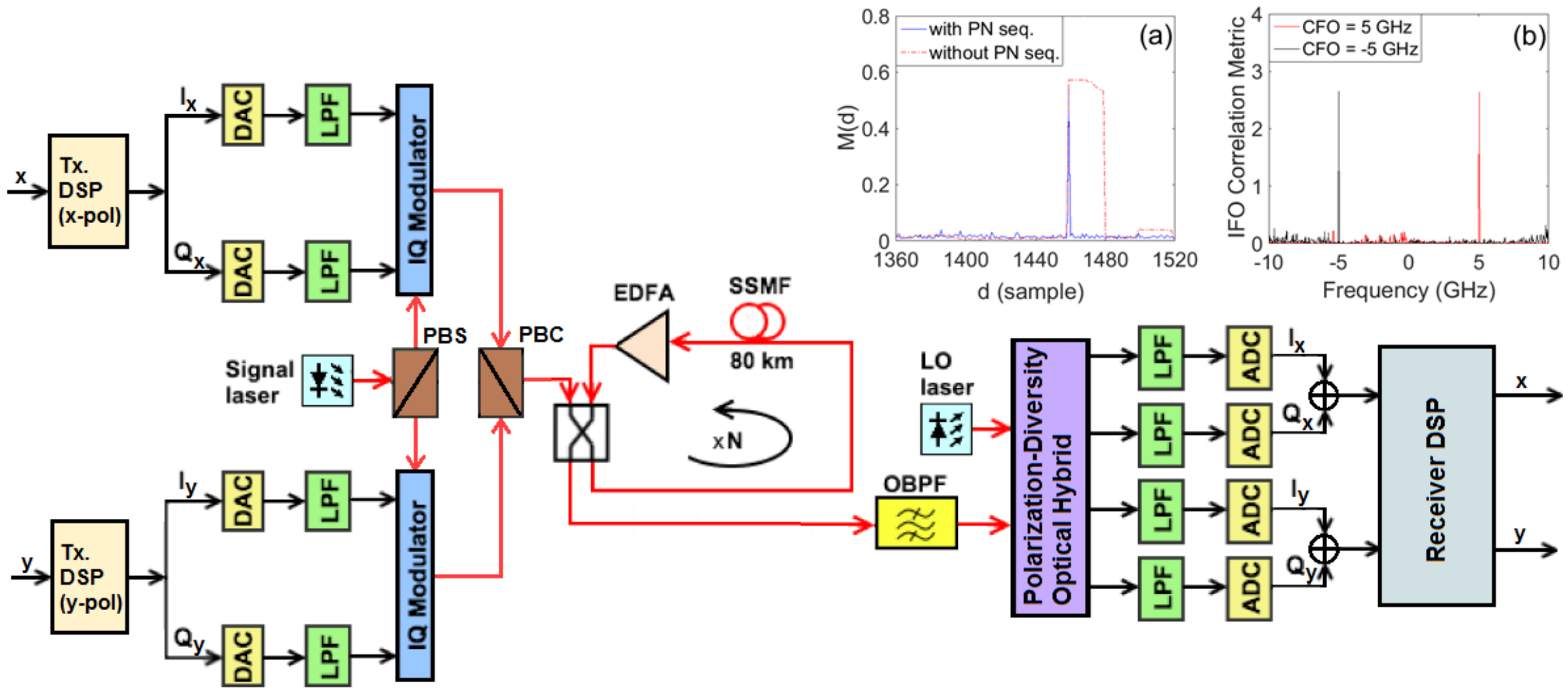

Fig. 2. Simulation setup. DAC: digital-to-analog converter. LPF: low-pass filter. PBS: polarization beam splitter. PBC: polarization beam coupler. EDFA: Erbium-doped fiber amplifier. SSMF: standard single-mode fiber. OBPF: optical band-pass filter. LO: local oscillator. ADC: analog-to-digital converter. Inset (a) Timing metric for *x*-pol, for a 20-sample relative delay, and a 5-GHz CFO. Inset (b) CFO correlation metric for -5 and 5 GHz CFOs.

## III. Simulation Setup And Results

To investigate the performance of the proposed scheme, a system model, whose schematic is depicted in Fig. 2, is built using VPI TransmissionMaker. The digital signal processing at the transmitter and receiver is performed in MATLAB. At the transmitter, for each polarization branch, a pseudo-random bit sequence is mapped onto 416 OFDM data subcarriers with 16-QAM modulation, which together with one unmodulated DC subcarrier, 10 unmodulated subcarriers around DC (for the RF-pilot method [14]), and 85 unmodulated edge subcarriers, are converted to the time-domain using a 512-point IFFT with a 46-sample CP. The OFDM signals from each polarization are fed to 40-GSa/s digital-to-analog converters, and then used to drive two null-biased I/Q modulators. A continuous wave laser with a linewidth of 100 kHz, average output power of 0 dBm, and center emission wavelength of 1550 nm is used as the signal laser. The optical signal is launched into a transmission link consisting of 10 spans of standard single-mode fiber (SSMF), with 80 km per span, and a 16-dB gain erbium-doped fiber amplifier. At the receiver, the signal is detected with a polarization-diversity optical hybrid. A laser with a linewidth of 100 kHz (and maximum CFO of 5 GHz) is used as the local oscillator, and the coherently-detected signal is sampled by 40-GSa/s analog-to-digital converters. An overlapped frequency-domain equalizer is used for CD compensation, and the RF-pilot method [14] is used for the joint compensation of the laser phase noise and residual CFO.

Each OFDM frame contains 1000 data symbols, and two TS for the joint synchronization and TA-CE. Thus, the overhead introduced by the TS is 0.2%, while the overhead for the CP is ~9%. With an additional 20.5% overhead for a triple-concatenated forward error correction (FEC) code [15], the net data rate is obtained as 197.6 Gb/s {40 GSa/s × 8 × (416/512) × [1/(1.002 × 1.09 × 1.205)]} [14]. Excluding the FEC overhead, the raw data rate is 238.1 Gb/s. The total OFDM overhead introduced to the system is thus 9.2% (31.6% after coding). Our previously proposed method in [9], if extended for PDM transmission, would require a total overhead of 11.2% (34% after coding). Inset (a) of Fig. 2 shows the measured timing metrics of the proposed algorithm for a CFO of 5 GHz and an OSNR of 4 dB. It is seen that including $p(n)$ when computing (4)-(7) yields an impulse-shaped timing metric with a distinct peak at the correct start of the frame. Conversely, without $p(n)$, the timing metric has a trajectory plateau with a width equal to the length of the delay, resulting in some uncertainty as to the actual start of the frame. Inset (b) of Fig. 2 shows plots of the correlation metric in (11) for 800-km SSMF transmission. It is seen that the correct CFO maximizes the correlation of the transmitted and received TS.

The robustness of the proposed scheme is then assessed in the presence of linear fiber impairments including amplified spontaneous emission noise, PDL, residual CD, and first-order PMD, as shown in Fig. 3. In addition, the required OSNR to achieve a target bit error rate (BER) of 1.8e-2, supported with soft decision FEC [15], is obtained as a function of all the considered linear impairments. The simulation results in Fig. 3 have been obtained for 1000 trial runs for each parameter value. For the PMD assessment, the SSMF is replaced with a PMD emulator, and the input state of polarization into the PMD emulator is set to 45° offset with respect to the transmitted signal. This is done to ensure the performance results are obtained for a worst-case alignment.

Figs. 3(a) and (b) show that the estimator can tolerate OSNR values as low as 4 dB, and PDL up to 10 dB with no frame synchronization errors. Fig. 3(c) shows that residual CD up to 800 ps/nm can be tolerated with no frame synchronization errors, beyond which the timing estimation gradually gets more inaccurate as the level of residual CD increases. Fig. 3(d) shows that for $\beta$ = 8, DGD up to 200 ps can be tolerated with no frame synchronization errors. This is in agreement with the condition for the value of $\beta$ as stated previously in Section II. The maximum CFO observed in all cases is ~9 MHz. The phase impairment brought about by this residual CFO is compensated for by the RF-pilot method. Fig. 3(e) shows that the BER performance obtained using the proposed method is identical to the performance obtained when there is ideal CFO estimation.

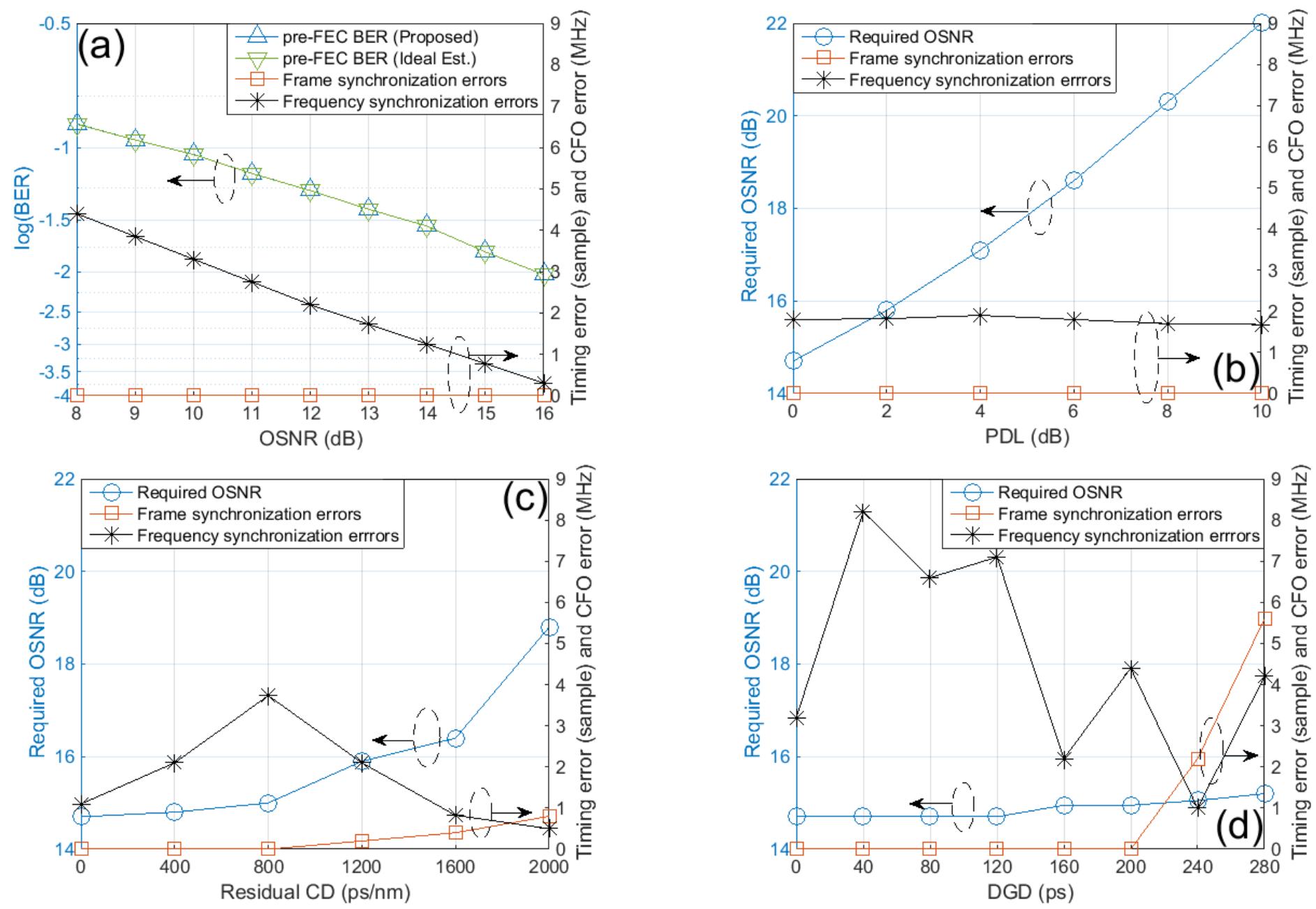

Fig. 3. (a) BER performance, frame and frequency synchronization performance in the presence of optical noise. (b)-(d) Required OSNR, frame and frequency synchronization performance in the presence of PDL ($\beta$=8), residual CD, and PMD, respectively.

With respect to the required OSNR in the presence of the considered impairments, the PDL tolerance curves in Fig. 3(b) show that the OSNR degrades rapidly with increasing PDL, with an OSNR penalty of 1 and 7.5 dB observed at 2-dB PDL and 10-dB PDL, respectively. Fig. 3(c) shows that TA-CE with negligible OSNR penalty is obtained for residual CD up to 800 ps/nm, beyond which the residual CD tolerance degrades. This tolerance can be improved by increasing the CP length, but this would come at the cost of increased overhead. Fig. 3(d) shows that a 0.25-dB OSNR penalty is obtained when the DGD is 200 ps. This OSNR penalty can be kept low by ensuring that the condition for the minimum value of $\beta$ is observed.

## IV. Conclusion

We have proposed a novel joint frame and frequency synchronization scheme for a 238.1 Gb/s (197.6-Gb/s after coding) PDM RGI-CO-OFDM system based on 16-QAM GCS arranged using the Alamouti scheduling scheme. The proposed scheme realizes a bandwidth-efficient solution because it uses the same training symbols for both the joint synchronization and TA-CE, resulting in a total overhead of 9.2% (31.6% after coding). We have shown that the proposed scheme has a wide CFO estimation range, as large as half the sample rate, and is robust against linear fiber impairments.